\documentclass[preprintnumbers,amsmath,amssymbm,preprint]{revtex4}
\usepackage{epsfig}
\usepackage{graphicx}

\begin{document}
%\title{Cloudy black holes: the large-mass limit}
%\title{Analytic treatment of cloudy black holes in the large-mass limit}
\title{The large-mass limit of cloudy black holes}
\author{Shahar Hod}
\affiliation{The Ruppin Academic Center, Emeq Hefer 40250, Israel}
\affiliation{ } \affiliation{The Hadassah Institute, Jerusalem
91010, Israel}
\date{\today}

\begin{abstract}
\ \ \ The interplay between black holes and fundamental fields has
attracted much attention over the years from both physicists and
mathematicians. In this paper we study {\it analytically} a physical
system which is composed of massive scalar fields linearly coupled
to a rapidly-rotating Kerr black hole. Using simple arguments, we
first show that the coupled black-hole-scalar-field system may
possess stationary bound-state resonances (stationary scalar
`clouds') in the bounded regime $1<\mu/m\Omega_{\text{H}}<\sqrt{2}$,
where $\mu$ and $m$ are respectively the mass and azimuthal harmonic
index of the field, and $\Omega_{\text{H}}$ is the angular velocity
of the black-hole horizon. We then show explicitly that these two
bounds on the dimensionless ratio $\mu/m\Omega_{\text{H}}$ can be
saturated in the asymptotic $m\to\infty$ limit. In particular, we
derive a remarkably simple analytical formula for the resonance mass
spectrum of the stationary bound-state scalar clouds in the regime
$M\mu\gg1$ of large field masses:
$\mu_n=\sqrt{2}m\Omega_{\text{H}}\big[1-{{\pi({\cal
R}+n)}\over{m|\ln\tau|}}\big]$, where $\tau$ is the dimensionless
temperature of the rapidly-rotating (near-extremal) black hole,
${\cal R}<1$ is a constant, and $n=0,1,2,...$ is the resonance
parameter. In addition, it is shown that, contrary to the flat-space
intuition, the effective lengths of the scalar field configurations
in the curved black-hole spacetime approach a {\it finite}
asymptotic value in the large mass $M\mu\gg1$ limit. In particular,
we prove that in the large mass limit, the characteristic length
scale of the scalar clouds scales linearly with the black-hole
temperature.
\end{abstract}
\bigskip
\maketitle

%]

\section{Introduction}

When we think about classical black holes, we usually imagine
spacetime regions which are bounded by one-way membranes that
irreversibly absorb all kinds of radiation and matter fields
\cite{Notefun}. This remarkable feature of the classical black-hole
horizon suggests that static matter configurations which are made of
fundamental fields cannot be supported in the spacetime region
exterior to the black-hole horizon. This physical expectation, first
discussed by Wheeler \cite{Whee,Car} more than four decades ago, is
known as the `no-hair' conjecture.

The no-hair conjecture \cite{Whee,Car}, which promoted the black
hole to the status of a fundamental object in general relativity
\cite{Noteatm}, has played a key role in the historical development
of black-hole physics. Early explorations of its regime of validity
have ruled out the existence of asymptotically flat static
configurations composed of a black hole nonlinearly coupled to
fundamental scalar fields \cite{Chas}, spinor fields \cite{Hart},
and massive vector fields \cite{BekVec}.

It should be emphasized, however, that the various `no hair'
theorems \cite{Chas,Hart,BekVec} do not rule out the possible
existence of {\it non}-static composed black-hole-fundamental-fields
configurations. In fact, it has recently \cite{Hodstat,Hodknc,Hodln}
been proved that linearized stationary scalar configurations
(bound-state scalar resonances) can be supported in the exterior
regions of {\it rotating} black holes \cite{Notelin}. In a very
interesting work, Herdeiro and Radu \cite{HerRa} have generalized
the results of \cite{Hodstat} to the nonlinear regime by numerically
solving the nonlinear coupled Einstein-scalar equations, thus
providing direct numerical evidence for the existence of these
non-static composed black-hole-scalar-field configurations.

The existence of non-static composed black-hole-bosonic-field
configurations \cite{Hodstat,HerRa,Notebos}, known as `cloudy black
holes'
%bosonic `clouds'
\cite{Notecloud}, is a direct consequence of
the superradiance phenomenon which characterizes the dynamics of
integer-spin fields in spinning black-hole spacetimes
\cite{Zel,PressTeu1,Ins1,Ins2}. In particular, the stationary
bound-state scalar configurations \cite{Hodstat,HerRa} are known to
be in resonance with the rotating black-hole horizon:
\begin{equation}\label{Eq1}
\omega_{\text{field}}=\omega_{\text{c}}\equiv m\Omega_{\text{H}}\ \
\ \ \text{with}\ \ \ \ m=1,2,3,...\  .
\end{equation}
where $\omega_{\text{field}}$ and $m$ are respectively the orbital
frequency of the scalar field and its azimuthal harmonic index, and
\cite{Noteunits}
\begin{equation}\label{Eq2}
\Omega_{\text{H}}={{a}\over{r^2_++a^2}}
\end{equation}
is the angular velocity of the black-hole horizon \cite{Notecrr}
(here $r_+$ and $a$ are respectively the horizon-radius and the
angular-momentum per unit mass of the black hole).

As shown in \cite{Zel,PressTeu1,Ins1,Ins2}, integer-spin modes which
are in resonance with the rotating black-hole horizon [that is,
bosonic modes which satisfy the resonance condition (\ref{Eq1})] do
not radiate energy into the central rotating black hole. This fact
suggests that, despite the inherent irreversibility associated with
the classical black-hole horizon, non-static bosonic field
configurations which co-rotate with the central spinning black hole
[see Eq. (\ref{Eq1})] may survive in the exterior black-hole
spacetime.

While the superradiance phenomenon \cite{Zel,PressTeu1,Ins1,Ins2}
insures that scalar fields co-rotating with the black hole with
$\omega_{\text{field}}=m\Omega_{\text{H}}$ are not swallowed by it,
an additional (and distinct) physical mechanism is required in order
to prevent the field from radiating its energy to infinity. For
massive fundamental fields, the mutual gravitational attraction
between the central black hole and the massive field provides the
required confinement mechanism \cite{Ins2}. In particular, for a
massive scalar field of mass $\mu$, field modes in the low frequency
regime \cite{Notedim}
\begin{equation}\label{Eq3}
\omega^2<\mu^2
\end{equation}
are prevented from escaping to infinity [that is, field modes in the
low frequency regime (\ref{Eq2}) decay exponentially fast at
asymptotically large distances from the central black hole, see Eq.
(\ref{Eq14}) below].

The main goal of the present study is to explore the physical
properties of the composed black-hole-scalar-field configurations
\cite{Hodstat,HerRa} (`cloudy' black holes \cite{Notecloud}) in the
regime
\begin{equation}\label{Eq4}
M\mu\gg1\
\end{equation}
of large field masses \cite{Notecom} (here $M$ is the mass of the
central spinning black hole). As we shall show below, these composed
cloudy black-hole configurations can be studied {\it analytically}
in the regime (\ref{Eq4}).

The paper is organized as follows: In Sec. II we describe the
composed Kerr-scalar-field system and formulate the characteristic
field equations which govern the dynamics of the scalar fields in
the rotating black-hole spacetime. In Sec. III we use simple
arguments in order to show that the coupled black-hole-scalar-field
system may possess stationary bound-state resonances in the bounded
regime $1<\mu/m\Omega_{\text{H}}<\sqrt{2}$. In Sec. IV we shall
derive the resonance condition which characterizes the stationary
bound-state resonances of the massive scalar fields in the
rapidly-rotating kerr black-hole spacetime. In particular, we shall
carefully determine the regime of validity of this resonance
condition. In Sec. V we shall solve the characteristic resonance
condition in the regime $M\mu\gg1$ of large field masses. In
particular, we shall obtain remarkably simple analytical formulas
which describe the resonance mass spectrum of the stationary
bound-state scalar clouds. In Sec. VI we shall explore the effective
spatial lengths of the bound-state scalar field configurations. In
particular, we shall show that, contrary to the flat-space
intuition, the effective lengths of the scalar clouds approach a
finite asymptotic value in the large mass $M\mu\gg1$ limit. We
conclude in Sec. VII with a brief summary of the main results.

\section{Description of the system}

The composed physical system we shall analyze consists of a scalar
field $\Psi$ of mass $\mu$ linearly coupled to a rapidly-spinning
Kerr black hole of mass $M$ and angular-momentum per unit mass $a$.
In the Boyer-Lindquist coordinate system $(t,r,\theta,\phi)$ the
black-hole spacetime is described by the line element
\cite{Chan,Kerr}
\begin{eqnarray}\label{Eq5}
ds^2=-{{\Delta}\over{\rho^2}}(dt-a\sin^2\theta
d\phi)^2+{{\rho^2}\over{\Delta}}dr^2+\rho^2
d\theta^2+{{\sin^2\theta}\over{\rho^2}}\big[a
dt-(r^2+a^2)d\phi\big]^2\  ,
\end{eqnarray}
where $\Delta\equiv r^2-2Mr+a^2$ and $\rho\equiv
r^2+a^2\cos^2\theta$. The black-hole (event and inner) horizons are
determined by the zeroes of $\Delta$:
\begin{equation}\label{Eq6}
r_{\pm}=M\pm(M^2-a^2)^{1/2}\  .
\end{equation}

The dynamics of the linearized massive scalar field $\Psi$ in the
curved black-hole spacetime is governed by the Klein-Gordon
(Teukolsky) wave equation \cite{Teuk,Stro}
\begin{equation}\label{Eq7}
(\nabla^\nu\nabla_{\nu}-\mu^2)\Psi=0\  .
\end{equation}
Substituting the field decomposition \cite{Notedec}
\begin{equation}\label{Eq8}
\Psi=\sum_{l,m}e^{im\phi}{S_{lm}}(\theta;a\omega){R_{lm}}(r;a,\omega)e^{-i\omega
t}\ ,
\end{equation}
into the characteristic Klein-Gordon wave equation (\ref{Eq7}), one
finds \cite{Teuk,Stro} that the angular wave function ${S_{lm}}$ and
the radial wave function ${R_{lm}}$ satisfy two coupled ordinary
differential equations [see Eqs. (\ref{Eq9}) and (\ref{Eq12}) below]
of the confluent Heun type \cite{Heun,Fiz1,Teuk,Abram,Stro,Hodasy}.

The angular wave functions $S_{lm}(\theta;a\omega)$, which are known
as the spheroidal harmonics, are determined by the angular equation
\cite{Heun,Fiz1,Teuk,Abram,Stro,Hodasy}
\begin{eqnarray}\label{Eq9}
{1\over {\sin\theta}}{{d}\over{\theta}}\Big(\sin\theta {{d
S_{lm}}\over{d\theta}}\Big) +\Big[K_{lm}+a^2(\mu^2-\omega^2)
-a^2(\mu^2-\omega^2)\cos^2\theta-{{m^2}\over{\sin^2\theta}}\Big]S_{lm}=0\
.
\end{eqnarray}
The regularity conditions at the poles (boundaries) $\theta=0$ and
$\theta=\pi$ single out a discrete set of angular eigenvalues
$\{K_{lm}\}$ which are labeled by the integers $l$ and $m$ with
$l\geq m$ \cite{Abram}. In the present study we shall analyze the
physical properties of the bound-state scalar resonances in the
eikonal (large-mass) regime
\begin{equation}\label{Eq10}
l=m\gg1\  ,
%\ \ \ \text{and}\ \ \ \ a\sqrt{\mu^2-\omega^2}\gg 1\  ,
\end{equation}
in which case the angular eigenvalues of (\ref{Eq9}) are given by
the simple asymptotic relation \cite{Yang,Notetbp}
\begin{equation}\label{Eq11}
K_{mm}=m^2-a^2(\mu^2-\omega^2)+O(m)\  .
\end{equation}

The radial Klein-Gordon (Teukolsky) equation is given by
\cite{Teuk,Stro}
\begin{equation}\label{Eq12}
{{d}\over{dr}}\Big(\Delta{{dR}\over{dr}}\Big)+UR=0\  ,
\end{equation}
where
\begin{equation}\label{Eq13}
U\equiv
\Delta^{-1}[(r^2+a^2)\omega-ma]^2+2ma\omega-K_{lm}-\mu^2(r^2+a^2)\ .
\end{equation}
Note that the angular eigenvalues
$\{{K_{lm}}(a\sqrt{\mu^2-\omega^2})\}$ couple the radial field
equation (\ref{Eq12}) to the characteristic spheroidal equation
(\ref{Eq9}) \cite{Notebr}.

The bound-state scalar configurations are characterized by radial
eigenfunctions which decay exponentially fast at asymptotically
large distances from the central black hole \cite{Ins2}:
\begin{equation}\label{Eq14}
R(r\to\infty)\sim {{1}\over{r}}e^{-\sqrt{\mu^2-\omega^2}r}\  ,
\end{equation}
where $\omega^2<\mu^2$ [see Eq. (\ref{Eq3})]. In addition, we shall
impose the physically motivated boundary condition of purely ingoing
waves (as measured by a comoving observer) at the outer horizon of
the black hole \cite{Ins2}:
\begin{equation}\label{Eq15}
R(r\to r_+)\sim e^{-i(\omega-\omega_{c})y}\  ,
\end{equation}
where the critical field frequency $\omega_{\text{c}}$ is determined
in (\ref{Eq1}) and the ``tortoise" radial coordinate $y$ is defined
by the relation $dy/dr=(r^2+a^2)/\Delta$ \cite{Noteho}.

The two boundary conditions (\ref{Eq14}) and (\ref{Eq15}), along
with the requirement (\ref{Eq1}), single out a discrete spectrum of
eigen field-masses which characterize the stationary bound-state
field configurations in the rotating black-hole spacetime. Below we
shall determine analytically this discrete resonance spectrum in the
regime $M\mu\gg1$ of large field masses.

\section{Upper and lower bounds on the allowed field masses of the stationary scalar clouds}

In the present section we shall use simple arguments in order to
obtain upper and lower bounds on the allowed field masses of the
stationary bound-state scalar clouds. To that end, we shall first
define the new radial function \cite{Hodbou}
\begin{equation}\label{Eq16}
\psi\equiv \Delta^{1/2}R\  ,
\end{equation}
in terms of which the radial Teukolsky (Klein-Gordon) equation
(\ref{Eq12}) takes the form
\begin{equation}\label{Eq17}
{{d^2\psi}\over{dr^2}}+(\omega^2_{\text{c}}-V)\psi=0\
\end{equation}
of a Schr\"odinger-like wave equation, where
\begin{equation}\label{Eq18}
\omega^2_{\text{c}}-V={{\Delta U+M^2-a^2}\over{\Delta^2}}\  .
\end{equation}
The asymptotic behavior of the effective binding potential $V(r)$ is
given by
\begin{equation}\label{Eq19}
V(r)=\mu^2-{{4M\omega^2_{\text{c}}-2M\mu^2}\over{r}}+O\Big({{1}\over{r^2}}\Big)\
.
\end{equation}

As discussed in the Introduction, the stationary scalar clouds
correspond to bound-state massive scalar resonances which are
trapped by the effective binding potential in the black-hole
exterior region. For the effective potential (\ref{Eq18}) to have a
trapping well, its asymptotic gradient must be positive:
\begin{equation}\label{Eq20}
V'\to 0^+\ \ \ \text{as}\ \ \ r\to\infty\  .
\end{equation}
Taking cognizance of Eqs. (\ref{Eq1}), (\ref{Eq3}), (\ref{Eq19}),
and (\ref{Eq20}), one finds that the regime of existence of the
stationary bound-state scalar clouds is bounded by
\begin{equation}\label{Eq21}
{{1}\over{\sqrt{2}}}\mu<m\Omega_{\text{H}}<\mu\  ,
\end{equation}
or equivalently
\begin{equation}\label{Eq22}
1<{{\mu}\over{m\Omega_{\text{H}}}}<\sqrt{2}\ .
\end{equation}
Note that the inequality (\ref{Eq22}) is compatible with the large
mass regime (\ref{Eq4}) for large values of the azimuthal harmonic
index $m$ [see Eq. (\ref{Eq10})].

Below we shall show explicitly that these two bounds on the
dimensionless ratio $\mu/m\Omega_{\text{H}}$ can be approached
arbitrarily close in the asymptotic $M\mu\gg1$ limit of large field
masses.

\section{The resonance condition and its regime of validity}

In the present section we shall obtain the characteristic resonance
condition for the stationary bound-state resonances of the massive
scalar fields in the rapidly-rotating (near-extremal) Kerr
black-hole spacetime. These stationary scalar resonances (scalar
clouds) correspond to the threshold (critical) frequency
\begin{equation}\label{Eq23}
\omega=\omega_{\text{c}}
\end{equation}
for superradiant scattering of bosonic fields in the rotating
black-hole spacetime [see Eq. (\ref{Eq1})].

The resonance condition for stationary scalar clouds in the regime
$M\mu=O(1)$ [which corresponds to the regime $m=O(1)$] was first
derived in \cite{Hodstat}. Here we shall generalize the analysis of
\cite{Hodstat} to the regime $M\mu\gg1$ of large scalar masses [see
Eq. (\ref{Eq4})]. In addition, we shall carefully determine the
regime of validity of the obtained resonance condition [see Eqs.
(\ref{Eq50}) and (\ref{Eq56}) below].

Before proceeding, it is worth emphasizing that extra care should be
taken in determining the exact regime of validity of the resonance
condition in the present analysis, which explores the physical
properties of the stationary bound-state scalar clouds in the {\it
double} asymptotic regime
\begin{equation}\label{Eq24}
m\gg1 \ \ \ \text{with}\ \ \ \tau\equiv{{r_+-r_-}\over{r_+}}\ll1\  .
\end{equation}
In particular, as we shall show below, it is highly important to
determine the exact asymptotic behavior of the combination $m\tau$
in the double limit (\ref{Eq24}).

It is convenient to define the dimensionless variables
\cite{Teuk,Stro}
\begin{equation}\label{Eq25}
x\equiv {{r-r_+}\over {r_+}}\ \ \ ;\ \ \ k\equiv 2\omega_{\text{c}}
r_+\ ,
\end{equation}
in terms of which the radial Teukolsky equation (\ref{Eq12}) becomes
\begin{equation}\label{Eq26}
x(x+\tau){{d^2R}\over{dx^2}}+(2x+\tau){{dR}\over{dx}}+UR=0\  ,
\end{equation}
where
\begin{equation}\label{Eq27}
U={{(1+x/2)^2k^2x}\over{x+\tau}}-K+2ma\omega_{\text{c}}-\mu^2[r^2_+(1+x)^2+a^2]\
.
\end{equation}

We shall first study the behavior of the radial function in the
near-horizon region
\begin{equation}\label{Eq28}
x\ll 1\  .
\end{equation}
In this region the radial equation is given by (\ref{Eq26}) with the
effective near-horizon potential $U\to U_{\text{near}}\equiv
k^2x/(x+\tau)-K+2ma\omega_{\text{c}}-\mu^2(r^2_++a^2)$. The physical
solution [that is, the one satisfying the ingoing boundary condition
(\ref{Eq15}) at the black-hole horizon] of Eq. (\ref{Eq26}) is given
by \cite{Morse,Abram}
\begin{equation}\label{Eq29}
R(x)=\Big({x\over \tau}+1\Big)^{-ik}{_2F_1}({1\over
2}+i\delta-ik,{1\over 2}-i\delta-ik;1;-x/\tau)\  ,
\end{equation}
where $_2F_1(a,b;c;z)$ is the hypergeometric function \cite{Abram}
and
\begin{equation}\label{Eq30}
\delta^2\equiv -K-{1\over
4}+2ma\omega_{\text{c}}+k^2-\mu^2(r^2_++a^2)\ .
\end{equation}
We shall henceforth consider the case of real $\delta$
\cite{Notedel,Noteel}.

Using Eq. 15.3.7 of \cite{Abram}, one can write the radial solution
(\ref{Eq29}) in the form
\begin{eqnarray}\label{Eq31}
R(x)&=&\Big({x\over \tau}+1\Big)^{-ik}\Big[
{{\Gamma(2i\delta)}\over{\Gamma({1/2}+i\delta-ik)\Gamma({1/2}+i\delta+ik)}}\Big({{x}\over{\tau}}\Big)^{-1/2+i\delta+ik}
\nonumber \\&& \times {_2F_1}({1\over 2}-i\delta-ik,{1\over
2}-i\delta-ik;1-2i\delta;-\tau/x)+(\delta\to -\delta)\Big]\ .
\end{eqnarray}
The notation $(\delta\to -\delta)$ in (\ref{Eq31}) means ``replace
$\delta$ by $-\delta$ in the preceding term." In the region
\begin{equation}\label{Eq32}
{{m\tau}\over{x}}\ll1\  ,
\end{equation}
one may use the property (see Eq. 15.1.1 of \cite{Abram})
\begin{equation}\label{Eq33}
_2F_1(a,b;c;z)\to 1\ \ \ \text{for}\ \ \ {{ab}\over{c}}\cdot z\to 0\
\end{equation}
of the hypergeometric function in order to approximate the radial
solution (\ref{Eq31}) by
\begin{eqnarray}\label{Eq34}
R(x)={{\Gamma(2i\delta)}\over{\Gamma({1/2}+i\delta-ik)\Gamma({1/2}+i\delta+ik)}}\Big({{x}\over{\tau}}\Big)^{-1/2+i\delta}
+(\delta\to -\delta)\ .
\end{eqnarray}
Taking cognizance of Eqs. (\ref{Eq28}) and (\ref{Eq32}), one
realizes that the expression (\ref{Eq34}) for the radial function is
valid in the range
\begin{equation}\label{Eq35}
m\tau\ll x\ll 1\  .
\end{equation}

We shall next study the behavior of the radial function in the
region
\begin{equation}\label{Eq36}
x\gg \tau\  ,
\end{equation}
in which case the radial Teukolsky equation (\ref{Eq26}) is well
approximated by
\begin{equation}\label{Eq37}
x^2{{d^2R}\over{dx^2}}+2x{{dR}\over{dx}}+V_{\text{far}}R=0\  ,
\end{equation}
with the effective far-region potential $U\to
U_{\text{{far}}}=(k+kx/2)^2-K+2ma\omega_{\text{c}}-\mu^2[r^2_+(1+x)^2+a^2]$.
The solution of the far-region radial equation (\ref{Eq37}) is given
by \cite{Morse,Abram}:
\begin{equation}\label{Eq38}
R(x)=N_1\times(2\epsilon)^{{1\over 2}+i\delta}x^{-{1\over
2}+i\delta}e^{-\epsilon x}{_1F_1}({1\over
2}+i\delta-\kappa,1+2i\delta,2\epsilon x)+N_2\times(\delta\to
-\delta)\ ,
\end{equation}
where $_1F_1(a,b,z)$ is the confluent hypergeometric function
\cite{Abram} and $\{N_1,N_2\}$ are normalization constants to be
determined below. Here we have defined
\begin{equation}\label{Eq39}
\epsilon\equiv \sqrt{\mu^2-\omega_{\text{c}}^2}r_+\
\end{equation}
and
\begin{equation}\label{Eq40}
\kappa\equiv {{k^2/2-(\mu r_+)^2}\over{\epsilon}}\ .
\end{equation}

In the region
\begin{equation}\label{Eq41}
mx\ll 1\  ,
\end{equation}
one may use the property (see Eq. 13.1.2 of \cite{Abram})
\begin{equation}\label{Eq42}
_1F_1(a,b,z)\to 1\ \ \ \text{for}\ \ \ {{a}\over{b}}\cdot z\to 0\
\end{equation}
of the confluent hypergeometric function in order to approximate the
radial solution (\ref{Eq38}) by
\begin{equation}\label{Eq43}
R(x)=N_1\times(2\epsilon)^{{1\over 2}+i\delta}x^{-{1\over
2}+i\delta}+N_2\times(\delta\to -\delta)\ .
\end{equation}
Taking cognizance of Eqs. (\ref{Eq36}) and (\ref{Eq41}), one
realizes that the expression (\ref{Eq43}) for the radial function is
valid in the range
\begin{equation}\label{Eq44}
\tau\ll x\ll m^{-1}\  .
\end{equation}

Taking cognizance of (\ref{Eq35}) and (\ref{Eq44}), one realizes
that, for rapidly-rotating (near-extremal) black holes with
$\tau\ll1$ [see Eq. (\ref{Eq24})], there is an overlap region
\begin{equation}\label{Eq45}
m\tau\ll x\ll m^{-1}\  ,
\end{equation}
in which {\it both} expressions for the radial eigenfunction [see
Eqs. (\ref{Eq34}) and (\ref{Eq43})] are valid. Matching the two
expressions (\ref{Eq34}) and (\ref{Eq43}) in the overlap region
(\ref{Eq45}), one finds
\begin{equation}\label{Eq46}
N_1(\delta)={{\Gamma(2i\delta)}\over{\Gamma({1\over
2}+i\delta-ik)\Gamma({1\over 2}+i\delta+ik)}}\tau^{{1\over
2}-i\delta}(2\epsilon)^{-{1\over 2}-i\delta}\ \ \ \ {\text{and}} \ \
\ \ N_2(\delta)=N_1(-\delta)\
\end{equation}
for the normalization constants of the radial function (\ref{Eq38}).

Using Eq. 13.5.1 of \cite{Abram}, one finds that the asymptotic
($x\to\infty$) behavior of the radial eigenfunction (\ref{Eq38}) is
given by \cite{Abram}
\begin{eqnarray}\label{Eq47}
R(x\to\infty)&\to&
\Big[N_1\times(2\epsilon)^{\kappa}{{\Gamma(1+2i\delta)}\over{\Gamma({1\over
2}+i\delta+\kappa)}}x^{-1+\kappa}(-1)^{-{1\over
2}-i\delta+\kappa}+N_2\times(\delta\to -\delta)\Big]e^{-\epsilon x}
\nonumber \\&& +
\Big[N_1\times(2\epsilon)^{-\kappa}{{\Gamma(1+2i\delta)}\over{\Gamma({1\over
2}+i\delta-\kappa)}}x^{-1-\kappa}+N_2\times(\delta\to
-\delta)\Big]e^{\epsilon x}\ .
\end{eqnarray}
The bound-state resonances of the massive scalar fields in the
black-hole spacetime are characterized by exponentially decaying
(bounded) eigenfunctions at spatial infinity [see Eqs. (\ref{Eq3})
and (\ref{Eq14})]. Thus, the coefficient of the asymptotically
exploding exponent $e^{\epsilon x}$ in (\ref{Eq47}) must vanish:
\begin{eqnarray}\label{Eq48}
N_1\times(2\epsilon)^{-\kappa}{{\Gamma(1+2i\delta)}\over{\Gamma({1\over
2}+i\delta-\kappa)}}x^{-1-\kappa}+N_2\times(\delta\to -\delta)=0\  .
\end{eqnarray}
Substituting $N_1$ and $N_2$ from (\ref{Eq46}) into (\ref{Eq48}),
one obtains the characteristic resonance condition
\begin{equation}\label{Eq49}
\Big[{{\Gamma(-2i\delta)}\over{\Gamma(2i\delta)}}\Big]^2{{\Gamma({1\over
2}+i\delta-ik)\Gamma({1\over 2}+i\delta+ik)\Gamma({1\over
2}+i\delta-\kappa)}\over{\Gamma({1\over 2}-i\delta-ik)\Gamma({1\over
2}-i\delta+ik)\Gamma({1\over
2}-i\delta-\kappa)}}\big(2\epsilon\tau\big)^{2i\delta}=1\
\end{equation}
for the bound-state resonances of the massive scalar fields in the
rapidly-rotating Kerr black-hole spacetime.

It is worth emphasizing again that, for large field masses
($M\mu\gg1$, or equivalently $m\gg1$), the resonance condition
(\ref{Eq49}) is valid in the asymptotic regime [see Eq.
(\ref{Eq45})]
\begin{equation}\label{Eq50}
\tau\ll m^{-2}
%\ll1\
\end{equation}
of near-extremal black holes.

In the regime $M\mu\gg1$ of large field masses (or equivalently, in
the regime $m\gg1$) one can use the following analytical
approximations for the various Gamma functions that appear in the
resonance condition (\ref{Eq49}) \cite{Abram}:
\begin{equation}\label{Eq51}
\Big[{{\Gamma(-2i\delta)}\over{\Gamma(2i\delta)}}\Big]^2=-\Big({{e}\over{2\delta}}\Big)^{8i\delta}[1+O(m^{-1})]\
,
\end{equation}
\begin{equation}\label{Eq52}
{{\Gamma({1\over 2}+i\delta-ik)}\over{\Gamma({1\over
2}-i\delta-ik)}}=e^{\pi\delta}e^{-2i\delta}(k+\delta)^{i(k+\delta)}(k-\delta)^{-i(k-\delta)}[1+O(m^{-1})]\
,
\end{equation}
\begin{equation}\label{Eq53}
{{\Gamma({1\over 2}+i\delta+ik)}\over{\Gamma({1\over
2}-i\delta+ik)}}=e^{-\pi\delta}e^{-2i\delta}(k+\delta)^{i(k+\delta)}(k-\delta)^{-i(k-\delta)}[1+O(m^{-1})]\
,
\end{equation}
and
\begin{equation}\label{Eq54}
{{\Gamma({1\over 2}+i\delta-\kappa)}\over{\Gamma({1\over
2}-i\delta-\kappa)}}=e^{-2i\delta}(\kappa^2+\delta^2)^{i\delta}e^{2i\kappa(\theta-\pi)}[1+O(m^{-1})]\
,\ \ \ \text{where}\ \ \ \theta\equiv\arctan(\delta/\kappa)\  .
\end{equation}
Substituting Eqs. (\ref{Eq51})-(\ref{Eq54}) into (\ref{Eq49}), one
obtains the resonance condition
\begin{equation}\label{Eq55}
-(k+\delta)^{2i(k+\delta)}(k-\delta)^{-2i(k-\delta)}
(\kappa^2+\delta^2)^{i\delta}e^{2i\kappa(\theta-\pi)}\Big({{e\epsilon\tau}\over{8\delta^4}}\Big)^{2i\delta}=1\
,
\end{equation}
which is valid in the asymptotic regime $M\mu\gg1$ (or equivalently,
in the asymptotic regime $m\gg1$). From Eq. (\ref{Eq55}) one finds
\cite{Notemis}
\begin{equation}\label{Eq56}
\tau{{e\epsilon}\over{8\delta}}(\alpha^2-1)\Big({{\alpha+1}\over{\alpha-1}}\Big)^{\alpha}(\gamma^2+1)^{1\over2}
e^{\gamma(\theta-\pi)}e^{\pi(n+{1\over2})/\delta}=1\  ,
\end{equation}
where
\begin{equation}\label{Eq57}
\alpha\equiv {{k}\over{\delta}}\ \ \ ; \ \ \
\gamma\equiv{{\kappa}\over{\delta}}\ ,
\end{equation}
and the resonance parameter $n$ is an integer.

In the next section we shall explicitly show that, the (rather
cumbersome) resonance equation (\ref{Eq56}) can be solved {\it
analytically} in the two physically interesting regimes: (1)
$\mu\lesssim\sqrt{2}\cdot m\Omega_{\text{H}}$, and (2) $\mu\gtrsim
m\Omega_{\text{H}}$. [Note that these two mass limits correspond to
the two boundaries of the regime of existence (\ref{Eq22}) of the
stationary bound-state scalar clouds].

\section{The stationary bound-state resonances of the composed Kerr-scalar-field system}

In the present section we shall derive (remarkably simple)
analytical formulas for the discrete mass spectrum, $\{\mu(m;n)\}$,
which characterizes the bound-state scalar resonances (the
stationary scalar clouds) in the rapidly-rotating Kerr black-hole
spacetime. In particular, we shall show below that, in the
large-mass regime $M\mu\gg1$, the bound-state resonances can almost
saturate the previously derived (upper and lower) bounds
(\ref{Eq22}) on the allowed field masses.

\subsection{The upper resonance regime of the mass spectrum}

We shall first consider bound-state scalar resonances in the
vicinity of the {\it upper} bound $\mu<\sqrt{2}m\Omega_{\text{H}}$
[see Eq. (\ref{Eq22})]. In particular, we shall write
$\mu=\sqrt{2}m\Omega_{\text{H}}(1-\Delta/2)$ with $\Delta\ll1$,
which implies \cite{Notewc}
\begin{equation}\label{Eq58}
M\mu={{m}\over{\sqrt{2}}}(1-\Delta/2)[1+O(\tau)]\ \ \ ; \ \ \
0\leq\Delta\ll1\ .
\end{equation}
We shall henceforth assume that $\Delta\gg\tau$ [see Eq.
(\ref{Eq65}) below]. Taking cognizance of Eqs. (\ref{Eq1}),
(\ref{Eq25}), (\ref{Eq30}), (\ref{Eq39}), (\ref{Eq40}),
(\ref{Eq54}), (\ref{Eq57}), and (\ref{Eq58}), one finds
\cite{Notecord}
\begin{equation}\label{Eq59}
\epsilon={1\over 2}m(1-\Delta)\ \ ; \ \ \delta={1\over
2}m(1+\Delta)\ \ ; \ \ \alpha=2(1-\Delta)\ \ ; \ \ \gamma=2\Delta\ \
; \ \ \theta={{\pi}\over{2}}-2\Delta\  .
\end{equation}

Substituting (\ref{Eq59}) into the resonance condition (\ref{Eq56}),
one obtains after some tedious (but straightforward) algebra
\cite{Notela}
\begin{equation}\label{Eq60}
\tau{{27e}\over{8}}\Big\{1-\Delta\Big[2(1+\ln3)+\pi\big(1+{{2n+1}\over{m}}\big)\Big]\Big\}e^{\pi(2n+1)/m}=1\
.
\end{equation}
Taking the logarithm of both sides of (\ref{Eq60}), one finds
\cite{Notela}
\begin{equation}\label{Eq61}
\Delta_n={{{{m}\over{2\pi}}\ln\big({{27e}\over{8}}\tau\big)+{1\over2}+n}
\over{{{m}\over{2\pi}}(2+2\ln3+\pi)+{1\over2}+n}}\
\ \ ; \ \ \ n\geq n_{\text{min}}\  ,
\end{equation}
where $n_{\text{min}}$ is the smallest integer $n$ for which
$\Delta_n\geq0$ \cite{Notekap} [or equivalently, the smallest
integer $n$ for which $m\ln(27e\tau/8)+\pi(1+2n)\geq 0$].

Denoting
\begin{equation}\label{Eq62}
{\cal R}\equiv
{{m}\over{2\pi}}\ln\big({{27e}\over{8}}\tau\big)+{1\over 2}-\left
\lfloor{{{m}\over{2\pi}}\ln\big({{27e}\over{8}}\tau\big)+{1\over
2}}\right \rfloor\ ,
\end{equation}
where $\left \lfloor{x}\right \rfloor$ is the floor function (the
largest integer less than or equal to $x$), one can express
(\ref{Eq61}) in the form
\begin{equation}\label{Eq63}
\Delta_n={{{\cal
R}+n}\over{{{m}\over{2\pi}}\big[2+2\ln3+\pi-\ln\big({{27e}\over{8}}\tau\big)\big]+{\cal
R}+n}}\ \ \ ; \ \ \ n=0,1,2,...\  .
\end{equation}
Note that ${\cal R}<1$, which implies \cite{Noteim}
\begin{equation}\label{Eq64}
\Delta_n<{{2\pi(1+n)}\over{m|\ln\tau|}}\ll1\ \ \ \text{for}\ \ \
n\ll m|\ln\tau|\  ,
\end{equation}
in accord with our assumption [see Eq. (\ref{Eq58})].

Note that in the extremal limit $\tau\to 0$ ($|\ln\tau|\gg1$) one
can approximate (\ref{Eq63}) by the remarkably compact expression
\cite{Notedt}
\begin{equation}\label{Eq65}
\Delta_n={{2\pi({\cal R}+n)}\over{m|\ln\tau|}}\ \ \ ; \ \ \
n=0,1,2,...\ \ (n\ll m|\ln\tau|)\ .
\end{equation}
Finally, taking cognizance of the relation (\ref{Eq58}), one finds
the simple analytical expression
\begin{equation}\label{Eq66}
M\mu_n={{m}\over{\sqrt{2}}}\Big[1-{{\pi({\cal
R}+n)}\over{m|\ln\tau|}}\Big]
%[1+O(\tau)]
\ \ \ ; \ \ \ n=0,1,2,...\ \ (n\ll m|\ln\tau|)\
\end{equation}
for the discrete family of field masses which characterize the
stationary bound-state resonances of the composed
black-hole-scalar-field system \cite{Noteemp1}.

\subsection{The lower resonance regime of the mass spectrum}

We shall next consider bound-state scalar resonances in the vicinity
of the {\it lower} bound $\mu>m\Omega_{\text{H}}$ [see Eq.
(\ref{Eq22})]. In particular, we shall write
$\mu=m\Omega_{\text{H}}(1+\nabla)$ with $\nabla\ll1$, which implies
\cite{Notewc}
\begin{equation}\label{Eq67}
M\mu={{m}\over{{2}}}(1+\nabla)[1+O(\tau)]\ \ \ ; \ \ \
0\leq\nabla\ll1\ .
\end{equation}
%We shall henceforth assume that $\nabla\gg\tau$ [see Eq. (\ref{Eq}) below].
Taking cognizance of Eqs. (\ref{Eq1}), (\ref{Eq25}), (\ref{Eq30}),
(\ref{Eq39}), (\ref{Eq40}), (\ref{Eq54}), (\ref{Eq57}), and
(\ref{Eq67}), one finds \cite{Notecord2}
\begin{equation}\label{Eq68}
\epsilon=m\sqrt{{{\nabla}/{2}}}\ \ ; \ \
\delta={{m}\over{\sqrt{2}}}(1-\nabla/2)\ \ ; \ \ \alpha=\sqrt{2}\ \
; \ \ \gamma={{1}\over{2\sqrt{\nabla}}}\ \ ; \ \
\theta=2\sqrt{\nabla}\ .
\end{equation}

Substituting (\ref{Eq68}) into the resonance condition (\ref{Eq56}),
one obtains after some tedious (but straightforward) algebra
\cite{Notelab}
\begin{equation}\label{Eq69}
\tau{{(\sqrt{2}+1)^{2\sqrt{2}}e^2}\over{16}}
e^{-\pi/2\sqrt{\nabla}}e^{\pi(2n+1)/\sqrt{2}m}=1\ .
\end{equation}
Taking the logarithm of both sides of (\ref{Eq69}), one finds that
$\nabla_n$ can be approximated by the remarkably compact expression
\begin{equation}\label{Eq70}
%\sqrt{\nabla_n}={{\pi}\over{{{\sqrt{2}\pi(2n+1)}\over{m}}+2\ln\Big[{{(\sqrt{2}+1)^{2\sqrt{2}}e^2}\over{16}}\tau\Big]}}
\nabla_n={{m^2}\over{2(2n+1)^2}}\ \ \ \text{for}\ \ \ n\gg
m|\ln\tau|\  .
\end{equation}
Note that $\nabla_n\ll1$ for $n\gg m$, in accord with our assumption
[see Eq. (\ref{Eq67})].

Finally, taking cognizance of the relation (\ref{Eq67}), one finds
the simple analytical expression
\begin{equation}\label{Eq71}
M\mu_n={{m}\over{{2}}}\Big[1+{1\over2}\Big({{m}\over{2n+1}}\Big)^2\Big][1+O(\tau)]
\ \ \ ; \ \ \ n\gg m|\ln\tau|\  .
\end{equation}
for the discrete family of field masses which characterize the
stationary bound-state resonances of the composed
black-hole-scalar-field system \cite{Noteemp2}.

\section{Effective lengths of the stationary bound-state scalar clouds}

A `no short hair' theorem for spherically-symmetric static black
holes was proved in \cite{Hod11,Notenun}. This theorem states that,
if a spherically-symmetric static black hole has hair, then this
hair (i.e. the external fields) must extend beyond the null circular
geodesic (the``photonsphere") of the corresponding black-hole
spacetime:
\begin{equation}\label{Eq72}
r_{\text{field}}\geq r_{\text{null}}\  .
\end{equation}

In the present section we shall test the validity of this `no short
hair' theorem \cite{Hod11} beyond the regime of spherically
symmetric static black holes. In particular, we shall now analyze
the effective lengths of the stationary bound-state scalar clouds. A
rough estimate for the characteristic heights of the scalar clouds
is given by the radial position $r=r_{\text{peak}}$ at which the
quantity $4\pi r^2|\Psi|^2$ attains its maximum value. In Table
\ref{Table1} we display the values of the dimensionless peak radii,
$x_{\text{peak}}(m)/\tau$, as directly obtained from the radial
eigenfunction (\ref{Eq29}) in the double asymptotic regime of
rapidly-rotating $\tau\ll1$ black holes and large $M\mu\gg1$ field
masses [see Eq. (\ref{Eq50})]. The data presented in Table
\ref{Table1} refer to bound-state scalar clouds which belong to the
upper resonance regime (\ref{Eq66}) of the mass spectrum
\cite{Notelonr}.

From Table \ref{Table1} one learns that the effective lengths of the
bound-state scalar clouds are a {\it decreasing} function of the
azimuthal harmonic index $m$ [or equivalently, a decreasing function
of the scalar field mass $\mu$, see Eq. (\ref{Eq66})]. In fact, one
finds that the data presented in Table \ref{Table1} is described
extremely well by the simple asymptotic formula:
\begin{equation}\label{Eq73}
{{x_{\text{peak}}}(\tau\to 0, l=m\gg1)\over{\tau}}\simeq
\alpha+{{\beta}\over{m}}+O(m^{-2})\ \ \ {\text{with}}\ \ \
\alpha\simeq 3.14 \ \ ; \ \ \beta\simeq 14.1\ .
\end{equation}

\begin{table}[htbp]
\centering
\begin{tabular}{|c|c|c|c|c|c|c|c|c|}
\hline
\ \ $l=m$\ \ \ & \ 10\ \ & \ 25\ \ & \ 50\ \ & \ 75\ \ & \ 100\ \ & \ 125\ \ \\
\hline \ \ $x_{\text{peak}}(m)/\tau$\ \ \ & \ \ 4.262\ \ \ & \ \
3.699\ \ \ & \ \ 3.439\ \ \ & \ \ 3.335\ \ \ & \ \ 3.276\ \ \ & \ \
3.237\ \ \ \\
\hline
\end{tabular}
\caption{Stationary bound-state resonances of the composed
black-hole-scalar-field system in the double asymptotic regime
$\tau\ll1$ with $l=m\gg 1$ [see Eq. (\ref{Eq50})]. We display the
dimensionless radius, $x_{\text{peak}}(m)/\tau$, corresponding to
the radial position $r=r_{\text{peak}}$ at which the quantity $4\pi
r^2|\Psi|^2$ attains its maximum value. One finds that the effective
heights of the scalar clouds, $x_{\text{peak}}(m)/\tau$, decrease
monotonically to an asymptotic {\it finite} value [see Eq.
(\ref{Eq74})] in the $m\gg1$ ($M\mu\gg1$) limit.} \label{Table1}
\end{table}

What we find most notable is the fact that the leading-order
coefficient $\alpha$ in (\ref{Eq73}) has a finite asymptotic value.
This fact suggests that the stationary bound-state scalar clouds are
characterized by the {\it finite} asymptotic ($M\mu\to\infty$) limit
\begin{equation}\label{Eq74}
{{x^{\infty}_{\text{peak}}}\over{\tau}}\simeq 3.14\
\end{equation}
of the dimensionless effective lengths, where
$x^{\infty}_{\text{peak}}\equiv x_{\text{peak}}(\tau\to 0,l=m\gg1)$.
At this point it is worth noting that the equatorial ($l=m\gg1$)
null circular geodesics of near-extremal ($\tau\to 0$) Kerr black
holes are characterized by the dimensionless ratio \cite{Bard}
\begin{equation}\label{Eq75}
{{x_{\text{null}}}\over{\tau}}={{2-\sqrt{3}}\over{2\sqrt{3}}}\simeq
0.077\  .
\end{equation}
Taking cognizance of Eqs. (\ref{Eq74}) and (\ref{Eq75}), one finds
\begin{equation}\label{Eq76}
x_{\text{peak}}(m)\geq x^{\infty}_{\text{peak}}>x_{\text{null}}\
\end{equation}
in the entire range of allowed field masses \cite{Notemaes}.

We find it remarkable that the composed black-hole-scalar-field
configurations conform to the lower bound (\ref{Eq72}) despite the
fact that they do {\it not} satisfy the main assumption (namely,
spherical symmetry) of the `no short hair' theorem presented in
\cite{Hod11}. This fact may suggest that the lower bound
(\ref{Eq72}) may be of general validity (that is, even beyond the
regime of spherically symmetric static black holes).

\section{Summary and discussion}

In summary, the physical properties of rapidly-rotating Kerr black
holes coupled to stationary bound-state scalar configurations
(linearized scalar `clouds') were studied {\it analytically}. To
that end, we have solved analytically the Kerr-Klein-Gordon
(Teukolsky) wave equation for linearized massive scalar fields in
the regime $M\mu\gg1$ of large field masses.

The main results obtained in this paper and their physical
implications are:

(1) It has been proved that, the regime of existence of the composed
black-hole-scalar-field configurations is bounded by
$1<\mu/m\Omega_{\text{H}}<\sqrt{2}$ [see Eq. (\ref{Eq22})].

(2) We have shown explicitly that the two (upper and lower) bounds
on the dimensionless ratio $\mu/m\Omega_{\text{H}}$ can be saturated
in the asymptotic (large-mass) limit $m\gg1$. In particular, we have
derived remarkably simple analytical formulas [see Eqs. (\ref{Eq66})
and (\ref{Eq71})] which describe the discrete resonance spectrum of
the stationary bound-state scalar clouds in the regime $M\mu\gg1$ of
large field masses.

(3) It is worth emphasizing that stationary bound-state scalar
clouds with $m=O(1)$ [or equivalently, with $M\mu=O(1)$] are {\it
weakly} bound to the central black hole in the sense that they are
characterized by the relation $M^2(\mu^2-\omega^2)\ll1$
\cite{Hodstat,HerRa}. Here, on the other hand, we have shown that
stationary scalar clouds in the regime $m\gg1$ are strongly bound to
the central black hole: in particular, it was shown that these
bound-state scalar clouds are characterized by the relation
$M^2(\mu^2-\omega^2)\gg1$.

(4) It has been shown that, contrary to the flat-space intuition,
the effective lengths of the stationary bound-state scalar clouds
approach a {\it finite} asymptotic value (which scales linearly with
the black-hole temperature) in the large mass $M\mu\gg1$ limit [see
Eq. (\ref{Eq74})]. In particular, we have shown that the non-trivial
spatial behavior of these scalar configurations must extend beyond
the equatorial null circular geodesic of the corresponding Kerr
black-hole spacetime [see Eqs. (\ref{Eq74}) and (\ref{Eq75})]
\cite{Notena,Zdr}.

We find this property of the ({\it non}-static, {\it
non}-spherically symmetric) scalar configurations quite remarkable
since the formal proof of the `no short hair' theorem (\ref{Eq72})
provided in \cite{Hod11} is restricted to the static sector of
spherically-symmetric black holes. To the best of our knowledge, the
fact that the composed black-hole-scalar-field configurations
conform to the lower bound (\ref{Eq72}) provides the first direct
evidence for a possible general validity \cite{Notegen} of the `no
short hair' property (\ref{Eq72}) for composed
black-hole-fundamental-fields configurations.

\bigskip
\noindent
{\bf ACKNOWLEDGMENTS}
\bigskip

This research is supported by the Carmel Science Foundation. I thank
C. A. R. Herdeiro and E. Radu for helpful correspondence. I would
also like to thank Yael Oren, Arbel M. Ongo and Ayelet B. Lata for
stimulating discussions.

%\newpage


\begin{thebibliography}{99}
\bibitem{Notefun} We refer here to fundamental fields which are not
associated with globally conserved charges.
\bibitem{Noteatm} By `fundamental object' we mean an object which is
characterized by only few conserved parameters, much like the
hydrogen atom in quantum mechanics.
\bibitem{Notelin} Since non-linear effects tend to stabilize an exterior black-hole hair,
we conjectured in \cite{Hodstat} the existence of non-static black
holes that are nonlinearly coupled to exterior scalar fields.
\bibitem{Notebos} Non-static composed black-hole-field configurations are expected to
exist for all types of integer-spin (bosonic) fields, see also
\cite{HerRa}. This expectation is based on the fact that the
superradiance phenomenon is a generic feature of the dynamics of all
bosonic fields in rotating black-hole spacetimes.
\bibitem{Notecloud} We shall use the term `cloudy black holes' to emphasize the fact that
these composed black-hole-bosonic-field configurations are made of
black holes which are linearly coupled to bosonic matter fields. On
the other hand, the term `hairy black holes' refers to black holes
which are coupled to nonlinear matter fields.
\bibitem{Noteunits} We use natural units in which $G=c=\hbar=1$.
\bibitem{Notecrr} Here $\omega_{\text{c}}\equiv m\Omega_{\text{H}}$
is the threshold (critical) frequency for superradiant scattering of
bosonic fields in the black-hole spacetime.
\bibitem{Notedim} Note that the field mass parameter $\mu$ stands for $\mu/\hbar$. Hence, it
has the dimensions of $($length$)^{-1}$.
\bibitem{Notecom} The regime (\ref{Eq4}) of large field masses
corresponds to scalar fields whose Compton wavelengths are small on
the scale set by the black-hole mass.
\bibitem{Notedec} Here $\omega$ is the conserved frequency of the wave field, and $(l,m)$ are
respectively the spheroidal harmonic index and the azimuthal
harmonic index of the field mode [see Eq. (\ref{Eq9}) below].
\bibitem{Notetbp} We shall also assume that
$a\sqrt{\mu^2-\omega^2}<m$, see Eq. (\ref{Eq66}) below.
\bibitem{Notebr} We shall henceforth omit the harmonic indexes $l$ and $m$ for brevity.
\bibitem{Notedel} One can choose $\delta>0$ without loss of generality.
\bibitem{Noteel} The case of imaginary $\delta$ values, which is technically much easier, was
studied in \cite{Hodstat}.
\bibitem{Notemis} Here we have used the relation $-1=e^{i\pi(2n+1)}$,
where the resonance parameter $n$ is an integer.
\bibitem{Notewc} Here we have used the relation $M\Omega_{\text{H}}={1\over 2}+O(\tau)$
for rapidly-rotating (near-extremal) Kerr black holes [see Eq.
(\ref{Eq2})].
\bibitem{Notecord} The correction terms in (\ref{Eq59}) are of order
$O(\Delta^2,\tau)$.
\bibitem{Notela} Here we have assumed the relation $n\Delta/m\ll1$, which is valid in the
regime $n\ll m|\ln\tau|^{1/2}$ [see Eq. (\ref{Eq65}) below].
\bibitem{Notekap} It is worth emphasizing that the validity of
(\ref{Eq59}) is restricted to the $\kappa\geq0$ regime (or
equivalently, to the $\Delta\geq 0$ regime).
\bibitem{Noteim} Here we have used the fact that
$2+2\ln3+\pi-\ln(27e/8)>0$.
\bibitem{Notedt} Note that our assumption $\Delta\gg\tau$ corresponds to the
requirement $m\tau|\ln\tau|\ll1$ [see Eq. (\ref{Eq65})]. Taking
cognizance of (\ref{Eq50}) and (\ref{Eq65}), one indeed finds
$\tau/\Delta\sim m\tau|\ln\tau|\ll \tau^{1/2}|\ln\tau|\to0$ in the
extremal $\tau\to 0$ limit.
\bibitem{Noteemp1} It is worth emphasizing again that the analytical
expression (\ref{Eq66}) is valid in the upper regime $\mu\lesssim
\sqrt{2} m\Omega_{\text{H}}$ of the mass spectrum [see Eq.
(\ref{Eq58})]. This implies $n\ll m|\ln\tau|$ in (\ref{Eq66}).
\bibitem{Notecord2} The correction terms in (\ref{Eq68}) are of order
$O(\nabla^2,\tau)$.
\bibitem{Notelab} Here we have assumed the relation $n\nabla/m\ll1$, which is valid in the
regime $n\gg m$ [see Eq. (\ref{Eq70}) below].
\bibitem{Noteemp2} It is worth emphasizing again that the analytical
expression (\ref{Eq71}) is valid in the lower regime $\mu\gtrsim
m\Omega_{\text{H}}$ of the mass spectrum [see Eq. (\ref{Eq67})].
This implies $n\gg m$ in (\ref{Eq71}).
\bibitem{Notena} It may be interesting to investigate the possible existence
of scalar clouds in the Kerr naked singularity spacetime \cite{Zdr}.
In particular, it would be interesting to explore the influence of
the trapped null geodesics in the Kerr naked singularity spacetime
\cite{Zdr} on the scalar clouds.
\bibitem{Notegen} That is, beyond the regime of spherically-symmetric static black
holes studied in \cite{Hod11}.
\end{thebibliography}
\end{document}